\documentclass[twocolumn,twoside,slac_two]{revtex4}
\usepackage{graphicx}
\usepackage{fancyhdr}
\usepackage{epstopdf}
\newcommand{\sNN}[1]{$\sqrt{s_{NN}} = #1$ GeV}
\begin{document}

\title{The Modification of high-$p_{T}$ hadro-chemistry in Au+Au collisions relative to p+p}

%

\author{A.R. Timmins for the STAR Collaboration}
\affiliation{Department of Physics and Astronomy, Wayne State University, Detroit, MI 48201, USA}

\begin{abstract}

We present high transverse momentum, $p_{T}$, pion ($\pi$), proton ($p$), kaon ($K$), and rho ($\rho$) spectra measured with the STAR experiment from p+p and Au+Au collisions with \sNN{200}. We find the $K/\pi$ ratio to be enhanced in Au+Au \sNN{200} collisions relative to p+p \sNN{200} collisions at $p_{T} >  5$ GeV/c. The enhancement persists until $p_{T} \sim 12$ GeV/c for central Au+Au 200 GeV collisions. We also show the nuclear modification factor, $R_{AA}$, measured at the same center of mass energy, and find $R_{AA}(K)$ and $R_{AA}(p)$ to be higher than $R_{AA}(\pi)$ at $p_T > 5$ GeV/c. Implications for medium induced modifications of jet chemistry is discussed.

\end{abstract}

\maketitle

\thispagestyle{fancy}


\section{Introduction}
Collisions of relativistic heavy-ions aim to create a unique state of matter, where quarks and gluons can move over large volumes in comparison to the typical size of a hadron. This state is known as the Quark Gluon Plasma, QGP \cite{STARWhite}. Measurements of jet quenching have provided strong evidence such a state is created in heavy-ion collisions at the Relativistic Heavy Ion Collider, RHIC \cite{JetQuench}. These measurements show a suppression of hadron production at high-$p_{T}$ relative to p+p collisions, scaled by the number binary collisions, and the disappearance of correlated back to back hadron pairs in certain kinematic regions. How the chemistry within the jet is altered in heavy-ion collisions may provide further information on the nature of jet quenching. In particular, there are two studies which predict jet chemistry will be significantly altered in the presence of a medium. The first, by Sapeta and Wiedemann \cite{SapWeid}, shows by increasing the in-medium parton splitting probabilities in the MLLA scheme, the $K/\pi$ and $p/\pi$ ratios may increase by a factor of 2 for in-medium jets. The increase in parton splitting probabilities simultaneously  leads to a suppression of hadron production in the high $z$ $(p_{T} (hadron) / p_{T} (jet))$ region of the fragmentation function relative to vacuum jets. The second study, by Liu and Fries \cite{JetConver}, shows that flavor conversions may alter the final state jet hadro-chemistry significantly. In this scheme, hard scattered partons can exchange flavor while traversing the medium resulting in $K$ production being less suppressed at high-$p_T$ relative to $\pi$ production, while $p$ production is similarly suppressed at high-$p_T$ relative to $\pi$ production. The suppression is characterized by the nuclear modification factor, $R_{AA}$ which is defined as:
\begin{equation}
R_{\rm{AA}}(p_{\rm T})\,=\,\frac{d^2N_{\rm{AA}}/dy dp_{\rm T}/\langle N_{\rm {bin}}\rangle}{d^2\sigma_{\rm{pp}}/dy dp_{\rm T}/\sigma_{\rm{pp}}^{\rm {inel}}}
\label{equ:RAA}
\end{equation}
where $\langle N_{bin} \rangle$ is the mean number of binary collisions occurring in heavy-ion collisions. In the absence of jet quenching, one naively expects $R_{AA}\sim 1$, while if jet quenching occurs in heavy-ion collisions, $R_{AA} < 1$ for the applicable $p_T$ region.

In these proceedings, we report precision measurements of inclusive hadron production at high-$p_T$ for a variety of particle species in p+p and Au+Au \sNN{200} collisions. These measurements extend previously published spectra \cite{starppPID, starAuAuPID} beyond $p_{T} \sim 6$ GeV/c. We first show the extended p+p spectra and compare our measurements to various Next to Leading Order calculations (NLO), in search of a suitable theoretical baseline to describe hadron production in p+p collisions. We then show the $K/\pi$ ratios in central heavy-ion collisions with a comparison to p+p, and present $R_{AA}$ for various particle species for central collisions. Finally, we show $R_{AA}$ as a function of system size for the $\pi$ and $K^{0}_{S}$ particles.


\section{Analysis}

\begin{figure*}[t]
\begin{tabular}{ll}
\includegraphics[width = 0.49\textwidth]{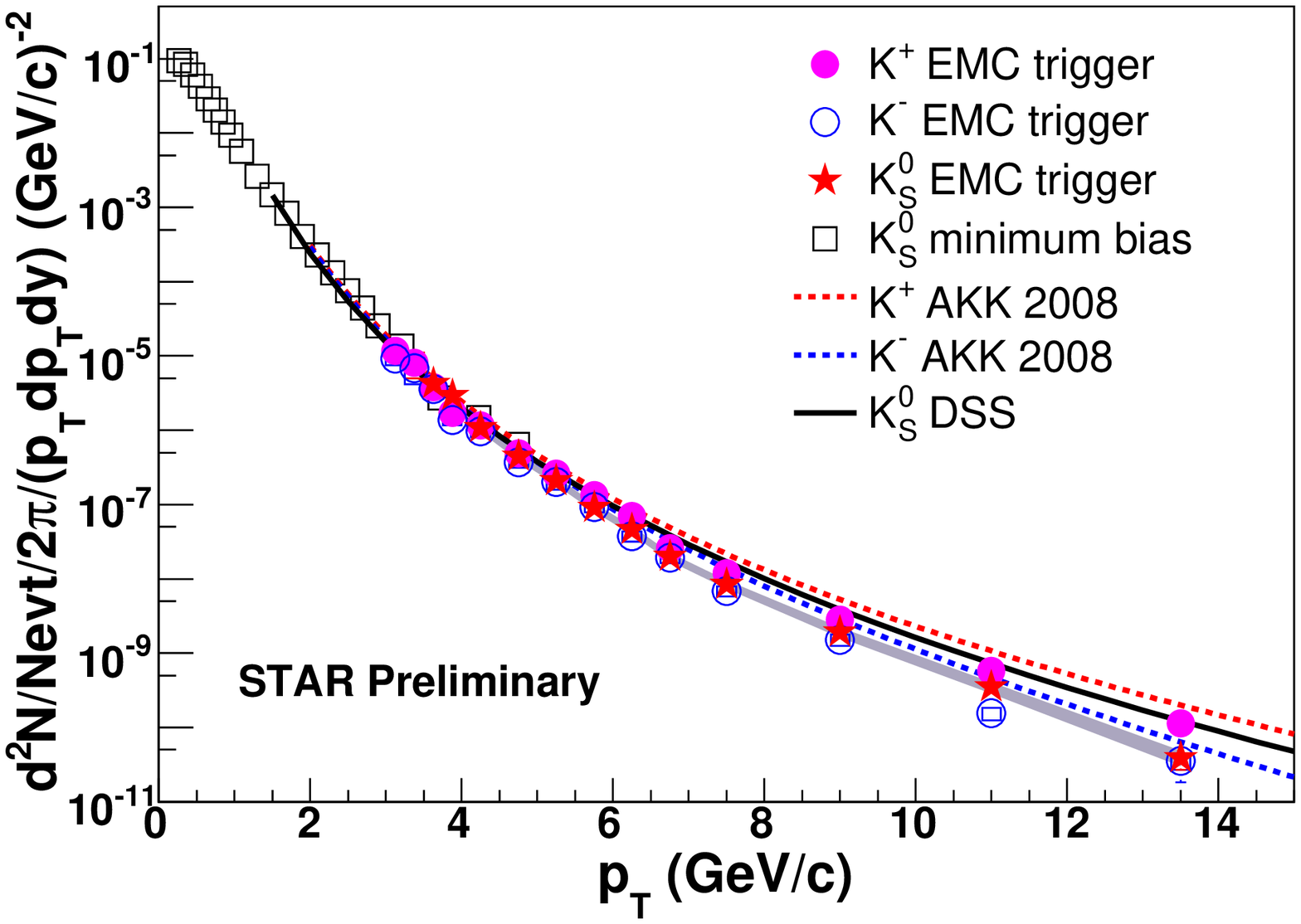}
&
\includegraphics[width = 0.5\textwidth]{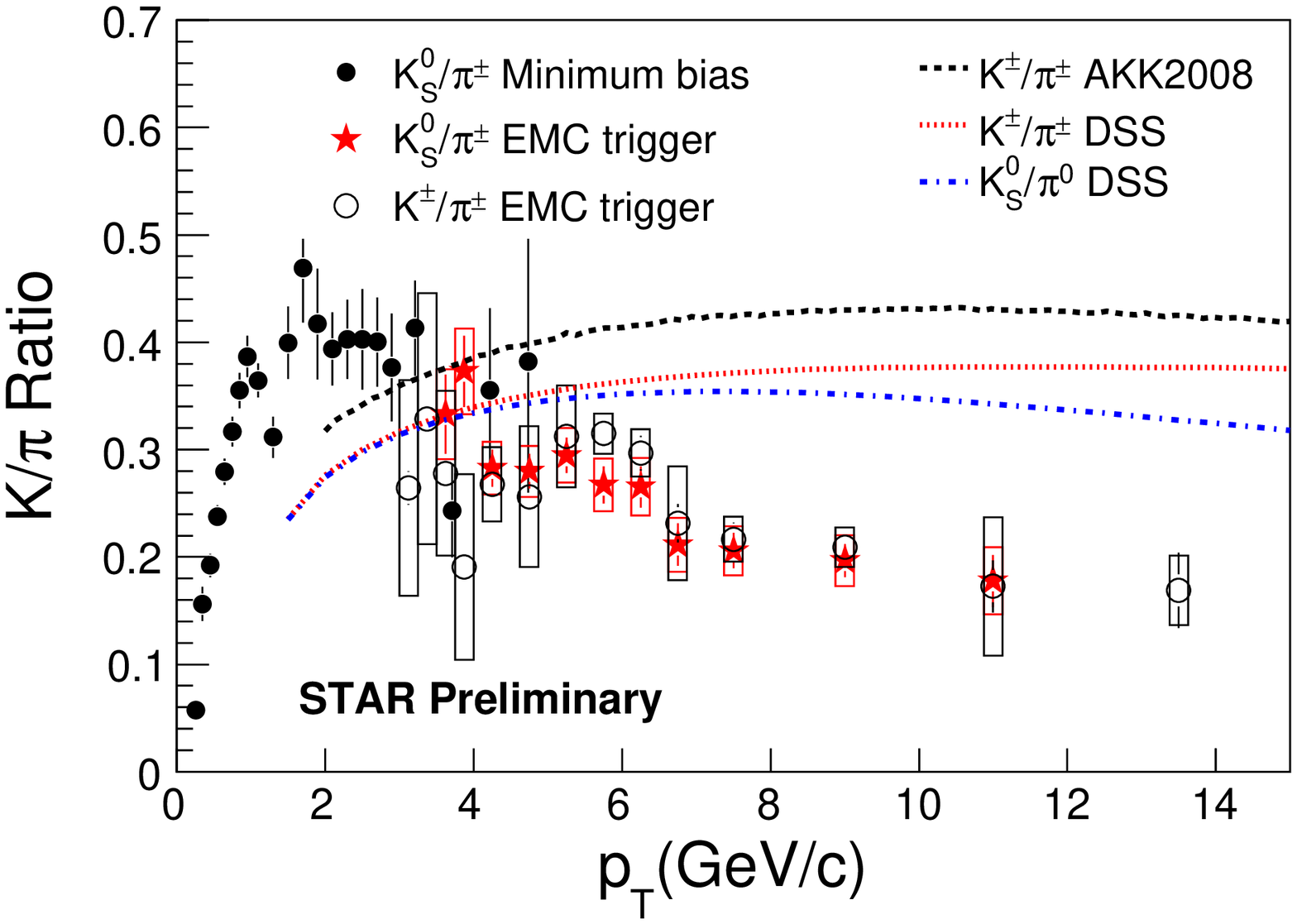}
\end{tabular}
\caption{Left Panel. Various mid-rapidity $K$ spectra from p+p \sNN{200} collisions. The open black symbols show previously published results \cite{starppPID}. The theory curves are described in the text. Right Panel. Mid-rapidity $K/\pi$ ratio as a function of $p_T$ in p+p \sNN{200}. collisions. The lines show statistical uncertainties, while the boxes show systematic uncertainties. The solid black symbols are previously published results \cite{starppPID}. EMC refers to the Electromagnetic Calorimeter.}
\label{Fig1}
\end{figure*}

The data presented from  p+p \sNN{200} collisions were collected in 2005. In order to maximize the reach of the p+p spectra, the data are from calorimeter triggered events which select events with high $p_{T}$ particles. Two trigger setups were used: jet-patch, which triggers on events where a large amount of neutral energy is deposited over many calorimeter towers, and high-tower, which triggers on events where a large amount of neutral energy is deposited in a single tower. Since the triggers only select a subset of events, trigger efficiency corrections are applied to present an unbiased invariant spectra. The STAR Time Projection Chamber, TPC, is used to detect charged particles. The $K^{\pm}$, $\pi^{\pm}$, and $p^{\pm}$ particles are identified via energy loss, $dE/dx$, measurements in the TPC, while the $K^0_{S}$ particles are identified via their weak decay products ($K^0_{S} \rightarrow \pi^{+}+\pi^{-}$).  The $\rho$ particles, like the $K^{0}_{S}$ particles, are also obtained via reconstruction of the respective decay products in the TPC. The charged and  $\rho$ particles were obtained from jet patch trigger events, while the $K^0_{S}$ particles were obtained from high tower events. All spectra are corrected for tracking inefficiencies and acceptance. The corrections are determined by embedding Monte Carlo particles into real events, then counting the number which are reconstructed under various cuts which are placed on the real data. Systematic uncertainties on the measurements are due to uncertainties in the trigger efficiency corrections, uncertainties in the TPC tracking efficiency/acceptance corrections, uncertainties in the raw yield extraction, and uncertainties due to the effects of finite momentum resolution. The Au+Au \sNN{200} data were collected in 2004. Two trigger setups were used; an online central trigger which obtained $\sim 20$ million 0-12\% central events, and a minbias trigger which obtained $\sim 20$ million Au+Au events independent of centrality. The techniques for extracting the various spectra are the same as for the p+p data. Uncertainties from the trigger efficiency corrections do not apply to the data from Au+Au collisions, since no trigger efficiency corrections are applied. 

\section{Results}

The left panel in figure \ref{Fig1} shows the extended mid-rapidity $K$ spectra in p+p \sNN{200} collisions. The open black symbols show previously published results \cite{starppPID}, and it is clear the new spectra are consistent within the overlapping $p_T$ range. We also show two NLO calculations; one from the Albino-Kniehl-Kramer (AKK) \cite{AKK2008}, the other from the DeFlorian-Sassot-Stratmann (DSS) \cite{DSS}. Although both sets of calculations appear to reproduce the shape of the various $K$ spectra, there are deviations at $p_{T} > 6$ GeV/c. Since both calculations use fits from measured $K$ fragmentation functions in $e^{+} + e^{-}$ collisions, these deviations maybe due to poor constraints on those fits from the data. The data in figure \ref{Fig1} in conjunction with independent measurements of $K$ fragmentation functions (at RHIC) should provide better constraints on models in the future. The right panel shows the mid-rapidity $K/\pi$ ratio as a function of $p_T$ in p+p \sNN{200} collisions. The solid black points show previously published data \cite{starppPID} and again we observe a consistency with the new data. We also observe both AKK and DSS again fail to describe the data at high-$p_{T}$. Clearly, an improved theoretically baseline is needed.

\begin{figure*}[t]
\begin{center}
\includegraphics[width = 0.9\textwidth]{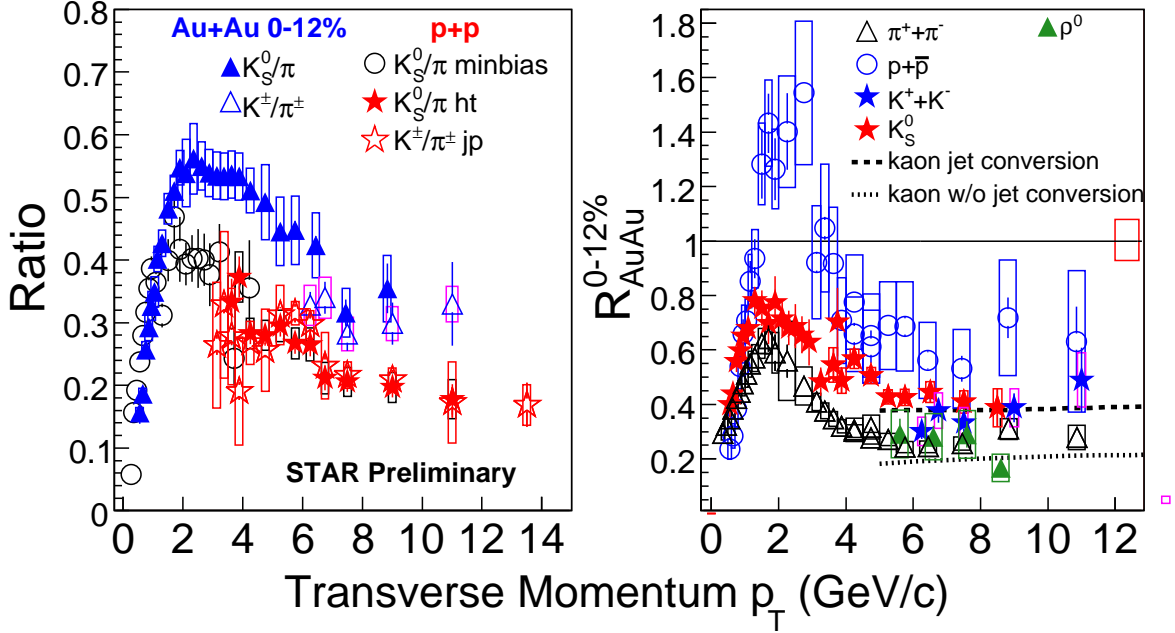}
\end{center}
\caption{Left Panel. The mid-rapidity $K/\pi$ ratio as a function of $p_{T}$ in p+p and central Au+Au 0-12\% \sNN{200} collisions. The acronyms jp (jet-patch) and ht (high-tower) refer to the type of calorimeter trigger used in p+p collisions and are further described in the text. The lines show statistical uncertainties, while the boxes show systematic uncertainties. Right Panel. Mid-rapidity nuclear modification factor, $R_{AA}$, for various particle species in Au+Au 0-12\% \sNN{200} collisions. Again, the lines show statistical uncertainties, while the boxes show systematic uncertainties.}
\label{Fig2}
\end{figure*} 

The left panel in figure \ref{Fig2} shows the mid-rapidity $K/\pi$ ratio as a function of $p_T$ in Au+Au $0-12\%$ \sNN{200} collisions. The p+p data are also shown for comparison. It is clear for $p_{T} > 2 $ GeV/c, the $K/\pi$ ratio is enhanced in heavy-ion collisions relative to p+p. In particular, the enhancement occurs at $p_T > 5$ GeV/c and persists until $p_{T} \sim 12$ GeV/c which is a clear sign jet-chemistry is indeed altered in heavy-ion collisions. This is qualitatively consistent with expectations from the Sapeta and Wiedemann model \cite{SapWeid}, where the $K/\pi$ ratio is roughly a factor of 2 higher in medium jets over all hadron $p_{T}$. It must be stated however the authors have yet to make specific predictions for RHIC inclusive spectrum measurements. Their current predictions are for LHC jets with energies above 50 GeV.

The right panel in figure \ref{Fig2} shows $R_{AA}$ for various particles species. The Au+Au centrality is again $0-12\%$ for \sNN{200} collisions. There are a number of things to note. Firstly, both $R_{AA}(K)$ and $R_{AA}(p)$ sit significantly above $R_{AA}(\pi)$ at $p_{T} > 5$ GeV/c. We speculate the underlying physics behind this does not relate to the higher mass of the $K$ or $p$ particles since we observe $R_{AA}(\rho)\sim R_{AA}(\pi)$. It has also been shown that $R_{AA}(\eta)\sim R_{AA}(\pi^0)$ in the same $p_{T}$ region \cite{etaRAA}. We also show predictions for the $K^{0}_{S}$ particles with and without the previously mentioned jet flavor conversions \cite{JetConver}. The $K$ data are quantitatively consistent with the conversion scenario. However, the same scheme predicts  $R_{AA}(p) \sim R_{AA}(\pi) $ which is quantitatively inconsistent with our data. Finally, it is trivial to show $R_{AA}(p) > R_{AA}(\pi)$ results from a higher $p/\pi$ in heavy-ion collisions relative to p+p. Again, this is qualitatively consistent with expectations in the Sapeta and Wiedemann model \cite{SapWeid}.

\begin{figure}[h]
\begin{center}
\includegraphics[width = 0.45\textwidth]{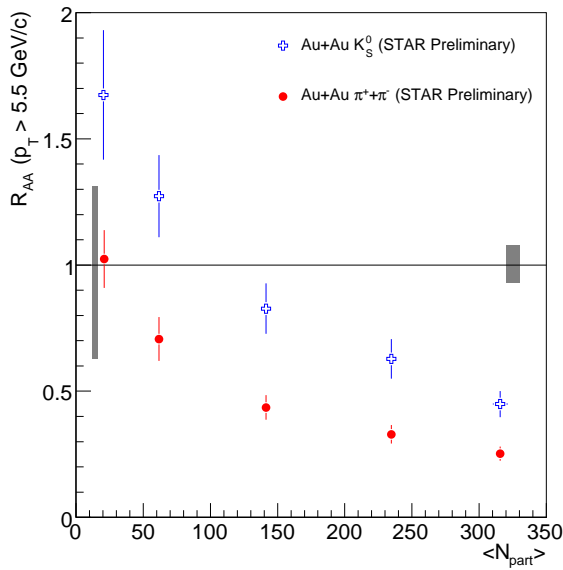}
\end{center}
\caption{Mid-rapidity integrated $R_{AA}(K^{0}_{S})$ and $R_{AA}(\pi)$ for $p_{T} > 5.5$ GeV/c in Au+Au \sNN{200} collisions. The uncertainties on the data points are statistical and systematic added in quadrature with respect to the yields. The left and right grey bands represent the typical uncertainties on $\langle N_{bin} \rangle $ for peripheral and central Au+Au \sNN{200} collisions respectively.}
\label{Fig3}
\end{figure} 

In figure \ref{Fig3}, we show the centrality dependance of integrated $R_{AA}(K^{0}_{S})$ and $R_{AA}(\pi)$ in Au+Au \sNN{200} collisions where $\langle p_{T} \rangle\sim 6.2$ GeV/c. As expected, we observe $R_{AA}(K^{0}_{S}) > R_{AA}(\pi)$ for central Au+Au collisions, however we also note this difference persists in 
peripheral collisions. The $K/\pi$ ratio is therefore higher in peripheral collisions relative to p+p at high-$p_{T}$. This raises the question whether jet chemistry is also altered in the smaller systems, or whether there is some other soft A+A production mechanism contributing in this $p_T$ range for all centralities. Measurements at higher $p_{T}$ from future higher statistics RHIC A+A runs may help in addressing these questions.


\section{Summary}

In summary, we have shown extended high-$p_{T}$ spectra for various particle species in p+p and Au+Au \sNN{200} collisions. We have found that the $K$ spectra are poorly described by current Next to Leading Order calculations for $p_{T} > 5$ GeV/c. We also present experimental evidence for modification of jet chemistry in heavy-ion collisions relative to p+p. This is characterized by a higher $K/\pi$ ratio for inclusive spectrum measurements at $p_{T} > 5 $ GeV/c in central Au+Au collisions, which persists until $p_{T} \sim 12$ GeV/c. The nuclear modification factor also displays a species dependance for central Au+Au collisions with $R_{AA}(p)$ and  $R_{AA}(K)$ being higher than $R_{AA}(\pi)$ at $p_{T} > 5$ GeV/c. We observe the integrated $R_{AA}(K)$ ($p_{T} > 5.5$ GeV/c) is higher than $R_{AA}(\pi)$ for peripheral collisions. Two models which predict changes in heavy-ion jet chemistry describe various aspects our data, however a complete quantitative description remains.



\bigskip 

\end{document}